\documentclass[letter, twocolumn]{jpsj3}
\usepackage{txfonts}
\usepackage{braket}
\usepackage{bm}

\usepackage{graphicx}
\usepackage{dcolumn}
\usepackage{bm}
\usepackage{amsmath}
\usepackage{times}
\usepackage[usenames]{color}

\title{Spin and Charge Fluctuations near Metal-Insulator Transition in Dimer-Type Molecular Solid}
\author{Naomichi Sato$^{1}$, Tsutomu Watanabe$^{2}$,  Makoto Naka$^{1}$,  and Sumio  Ishihara$^{1}$\thanks{ishihara@cmpt.phys.tohoku.ac.jp}}
\inst{$^1$Department of Physics, Tohoku University, Sendai 980-8578, Japan \\
$^2$  Department of Natural Science, Chiba Institute of Technology, Narashino, Chiba 275-0023, Japan
} 

\abst{
Spin and charge fluctuations at vicinity of metal-to-Mott insulator transitions are studied
in an organic solid with molecular dimers.  
The extended Hubbard model taking account of the internal electronic degree of freedom in a molecular dimer
is analyzed using the variational Monte Carlo method. 
Three kinds of the electronic phases, i.e. a metallic phase, an antiferromagnetic insulating phase and a polar charge ordered phase, compete with each other in the ground state.  
 It is found that the polar-charge fluctuation is dominant in a wide range of the molecular dimerization  and Coulomb interaction amplitudes, and is enhanced remarkably near the metal-insulator phase boundary, in which 
the spin fluctuation is almost unchanged. 
Implication for the $\kappa$-type BEDT-TTF salts is discussed. 
}

\begin{document}
\maketitle


Metal-to-insulator (MI) transition owing to electron correlation has been accepted as a fascinating topic in condensed matter physics since early 1900's~\cite{Imada}. 
Two routes to realize the MI transition are known: the carrier doping and the band-width control.  
A prototypical system for the former-type MI transition is the high transition-temperature superconducting (HTSC) cuprates.
By hole or electron doping due to the chemical substitution of cations, the N$\rm \acute e$el temperature for an antiferromagnetic (AFM) Mott insulating phase decreases gradually, and a superconductivity appears. 
It is widely believed that the AFM fluctuation resulted from a melting of the AFM order plays a crucial role on the emergence of the superconductivity. 

On the other hand, the MI transition induced by the band-width control is seen in molecular organic solids. 
By chemical substitution and/or applying the pressure, the band width is changed owing to their flexible framework, and the system undergoes the MI transition~\cite{Kanoda}. 
%
One of the well-known examples is a series of $\kappa$-(BEDT-TTF)$_2$X (X is an anion). 
The crystal structure consists of the alternating BEDT-TTF and X layers, and the dimers built by the BEDT-TTF molecules are arranged on an anisotropic triangular lattice. 
The number of holes per the BEDT-TTF molecular dimer is one, and the antibonding molecular orbitals form a half-filled band. 
Thus, the system is identified as a Mott insulator, termed a dimer-Mott (DM) insulator, 
in the presence of the strong intra-dimer electron-electron interactions~\cite{Kino, Seo}.  
An AFM Mott insulating phase is changed into the superconducting phase by substitution of X and/or applying pressure across the first-order phase transition boundary~\cite{Komatsu, Miyagawa}. 
The AFM fluctuation is expected to be plausible candidates for the attractive interaction in the superconductivity. 
In fact, the divergent increases of the AFM fluctuation toward the MI transition are of avail for the mechanisms of the superconductivity in the weak coupling theories~\cite{Kontani,Kuroki}. 

The recent discovered dielectric anomalies in $\kappa$-(BEDT-TTF)$_2$Cu$_2$(CN)$_3$ 
triggered reinvestigations of the electronic structure beyond the DM insulator picture~\cite{Jawad, Lang, Iguchi, Jawad2}. 
A possible interpretation of the electric dipole moments is the electronic charge distribution without the inversion symmetry inside the dimers~\cite{Drichko,Tomic,Kishida,Yakushi}. 
This scenario is supported by the optical spectra under the electric field~\cite{Kishida} 
and a broadening of the phonon spectra~\cite{Yakushi}. 
The charge ordered (CO) insulating state associated with the electric dipole moment and its competition with the DM insulating state were proposed by the theoretical studies~\cite{Naka, Hotta,Gomi,Dayal,Sekine}. 
Beyond the conventional MI transition, 
the spin and charge fluctuation as well as the superconductivity near the MI transition should be reexamined from the view point of the competition among the metallic phase, the DM insulating phase and the polar CO phase.  

In this Letter, 
the spin and charge fluctuations near the MI transition are studied as a subject of the dimer-type organic molecular solids. 
The extended Hubbard model with the internal electronic degree of freedom in molecular dimers is analyzed using the variational Monte Carlo (VMC) method\cite{Willkins}. 
The ground state phase diagram, and the spin and charge correlation functions near the MI transitions are calculated. 
We find that the polar charge fluctuation in the metallic phase is much larger than the spin fluctuation, and 
increases toward the first-order MI transition boundary. 
This characteristic is insensitive to the cluster size. 
An origin of the discrepancy between the spin and polar-charge fluctuations is discussed. 

\begin{figure}[t]
\begin{center}
\includegraphics[width=\columnwidth,clip]{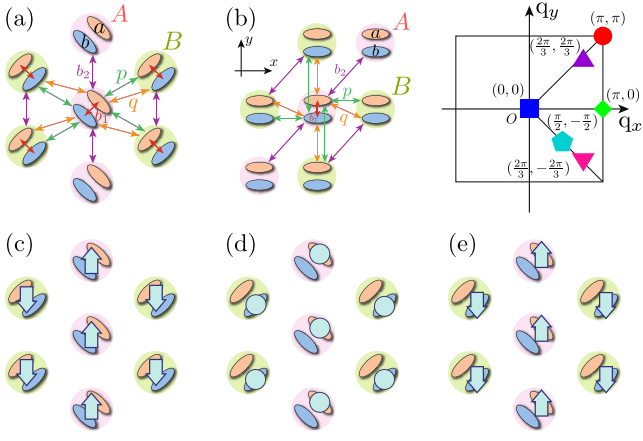}
\end{center}
\caption{(Color online) 
(a) A schematic lattice structure in the BEDT-TTF plane in $\kappa$-(BEDT-TTF)$_2$X. 
Ovals and circles represent the BEDT-TTF molecules and the molecular dimer units, respectively.  
Symbols $(A, B)$ and $(a, b)$ denote the two kinds of the molecular dimers, and the two molecules in a dimer unit, respectively. 
Arrows represent the electron transfer integrals and the electron-electron Coulomb interactions. 
(b) A square lattice in which the correlation functions are calculated, and the corresponding first Brillouin zone. 
%
(c) The AFM structure, (d) the CO structure and (e) the CO-AFM structure. 
Bold arrows and filled circles represent the spin directions, and the electrons inside of the dimers, respectively. 
}
\label{fig:model}
\end{figure}

The extended  Hubbard model with the internal electronic degree of freedom in the molecular dimers
consists of the two terms as 
\begin{align}
{\cal H}={\cal H}_{\rm intra}+{\cal H}_{\rm inter} . 
\label{eq:hamiltonian}
\end{align}
The first and second terms represent the intra- and inter-dimer parts, respectively, given as  
\begin{align}
{\cal H}_{\rm intra}&=
U\sum_{i  \mu} n_{i \mu \uparrow} n_{i \mu \downarrow} \nonumber \\
&+t_{b1}\sum_{i  \sigma} \left ( 
c_{i  a \sigma}^\dagger c_{i b \sigma} + {\rm H.c.} 
\right ) 
+
V_{b1} \sum_{i } n_{i a} n_{i b} , 
\label{eq:Hintra}
\end{align}
and 
\begin{align}
 {\cal H}_{\rm inter}&=
 \sum_{\langle i i' \rangle} \sum_{ \mu \mu'}\sum_\sigma 
t_{i i'}^{ \mu  \mu'} 
\left ( c_{i  \mu \sigma}^\dagger c_{i'  \mu' \sigma}
+{\rm H.c.} \right ) 
\nonumber \\
& +\sum_{ \langle ii' \rangle} 
\sum_{ \mu \mu'} V_{i i'}^{\mu  \mu'} 
 n_{i  \mu} n_{i' \mu'} . 
\label{eq:Hinter}
\end{align}
where  $c_{i  \mu \sigma}^\dagger$ ($c_{i \mu \sigma}$) is 
the creation (annihilation) operator for an electron at $i$-th dimer with molecule $\mu(=a, b)$ and spin $\sigma (=\uparrow, \downarrow)$. 
The number operator is defined as $n_{i \mu }=\sum_{\sigma} n_{i \mu \sigma}$ 
with $n_{i  \mu \sigma}=c_{i \mu \sigma}^\dagger c_{i \mu \sigma}$.  
The first term in Eq.~(\ref{eq:Hintra}) represents 
the electron-electron interaction inside a molecule.  
The 2nd and 3rd terms represent 
the electron hopping and the electron-electron interaction  between the molecules, respectively. 
The inter-dimer part in Eq.~(\ref{eq:Hinter}) represents the electron hoppings  
and the electron-electron interactions between $(i, \mu)$ and $(i',  \mu')$. 

Motivated from the $\kappa$-type BEDT-TTF salts, we introduce the two-dimensional arrangement of molecules and the notations for the bonds, in which the interactions are taken into account, as shown in Fig.~\ref{fig:model}(a).  
We adopt $t_p$, $t_{b1}$ and $t_{b2}$ for the hopping integrals, and $V_{p}$, $V_q$, $V_{b1}$ and $V_{b2}$ for the inter-molecule Coulomb interactions.~\cite{Komatsu} 
%
In the numerical calculation, we chose $t_{b2}/t_p=0.8$, $V_{b1}/t_p=6$, $V_{b2}/t_p=2.4$, $U/t_p=10$, 
and $(V_p+V_q)/(2t_p)=2.85$, and vary $t_{b1}$ and $\Delta V\equiv V_p-V_q$. 
The average electron number density is fixed to be $N_{\rm ele}/N=1$ where $N_{\rm ele}$ and $N$ are 
the numbers of the electrons and the molecular-dimers, respectively.

The variational wave function for the VMC method is represented as a product of the one-body wave function $\Phi$ and the many-body correlation factors such that 
\begin{align}
\Psi=P_G P_{V} P_Q \Phi . 
\end{align}
We introduce the three kinds of the correlation factors: 
the Gutzwiller factor 
$P_G=\prod_{ i \mu} 
\left (1-\eta n_{i \mu \uparrow}n_{ i \mu \downarrow} \right )$,  
the Jastraw factor  
$P_{V}=\prod_{ (i \mu, i' \mu')} 
\left (1-v_{(i \mu, i' \mu')} n_{i \mu}n_{i' \mu'} \right ) $, 
and the correlation factor for the doublon-holon bound state 
$P_Q=\prod_{(i \mu, i' \mu')}
\left ( 1-\mu_{(i \mu, i' \mu')} Q_{(i \mu, i'  \mu')} \right ) $~\cite{Yokoyama1}. 
Symbols $(i \mu, i' \mu')$ represents a neighboring molecule pair. 
We define 
$Q_{(i \mu, i' \mu')} =\left [ d_{i \mu} (1-e_{i' \mu'})+e_{i \mu} (1-d_{i' \mu'}) \right ]$
with the doublon operator 
$d_{i \mu}=n_{i \mu \uparrow}n_{i \mu \downarrow}$, 
and the holon operator 
$e_{i \mu}=(1-n_{i \mu \uparrow})(1-n_{i \mu \downarrow})$. 
We introduce the variational parameters $v_{p}$, $v_{q}$, $v_{b1}$, and $v_{b_2}$ for $v_{(i \mu, i' \mu')}$ in the $p$, $q$, $b1$, and $b2$ bonds, respectively, and $\mu_{p}$, $\mu_{b1}$, and $\mu_{b2}$ for $\mu_{(i \mu, i' \mu')}$ in the $p$, $b1$, and $b2$ bonds, respectively. 
Other components of $\nu_{(i \mu, i' \mu')}$ and $\mu_{(i \mu, i' \mu')}$ are set to be zero. 
The spin and charge ordered states are taken into account in $\Phi$, which is given by the Hartree-Fock approximation. 
We introduce the AFM structure and the polar CO structure, where the electron densities inside the dimer molecules are polarized, as shown in Figs.~\ref{fig:model}(c) and \ref{fig:model}(d), respectively. 
In $\Phi$, the mean fields characterizing them are optimized as variational parameters, $\Delta_{\rm AF}$ and $\Delta_{\rm CO}$.
A coexistence of the polar CO and AFM order shown in Fig.\ref{fig:model}(e) is also taken into account as a candidate state. 
It is useful to introduce the molecular configuration 
shown in Fig~\ref{fig:model}(b), in which the dimer units are arranged in a square lattice and 
the two molecules at the $i$-th dimer unit are assumed to be at the same site with a position ${\bf r}_i$. 
A cluster of the $6 \times 6$ sites with the periodic- and antiperiodic-boundary conditions along the $x$ and $y$ axes, respectively (see Fig.~\ref{fig:model}(b)), is adopted in most of the calculations, and that of the $12 \times 12$ sites is used to check the size dependence. 

The order parameters characterizing AFM and polar CO are defined as
\begin{align}
M_{\rm AFM} = 2N^{-1} \sum^{N}_{i} \langle S^{z}_{i} \rangle e^{i {\bf Q} \cdot {\bf r}_i}
\end{align}
and
\begin{align}
M_{\rm CO} = N^{-1} \sum^{N}_{i} \langle P_{i} \rangle  e^{i {\bf Q} \cdot {\bf r}_i},
\end{align}
respectively, where $S_i^z=(1/2)\sum_{\mu=(a,b)} 
(n_{i \mu \uparrow}-n_{i \mu \downarrow})$ is the spin operator and 
$P_i=\sum_{\sigma=(\uparrow, \downarrow)}(n_{i a \sigma}-n_{i b \sigma})$ is the electric polarization operator inside the molecular dimer. The AFM and polar CO structures shown in Figs.~\ref{fig:model}(c) and \ref{fig:model}(d) are characterized 
by the wave vector ${\bf Q} = \left( \pi, \pi \right)$.
The expectation values $\langle S^z_i \rangle$ and $\langle P_i \rangle$ are calculated by using the optimized wave functions.
\begin{figure}[t]
\begin{center}
\includegraphics[width=\columnwidth,clip]{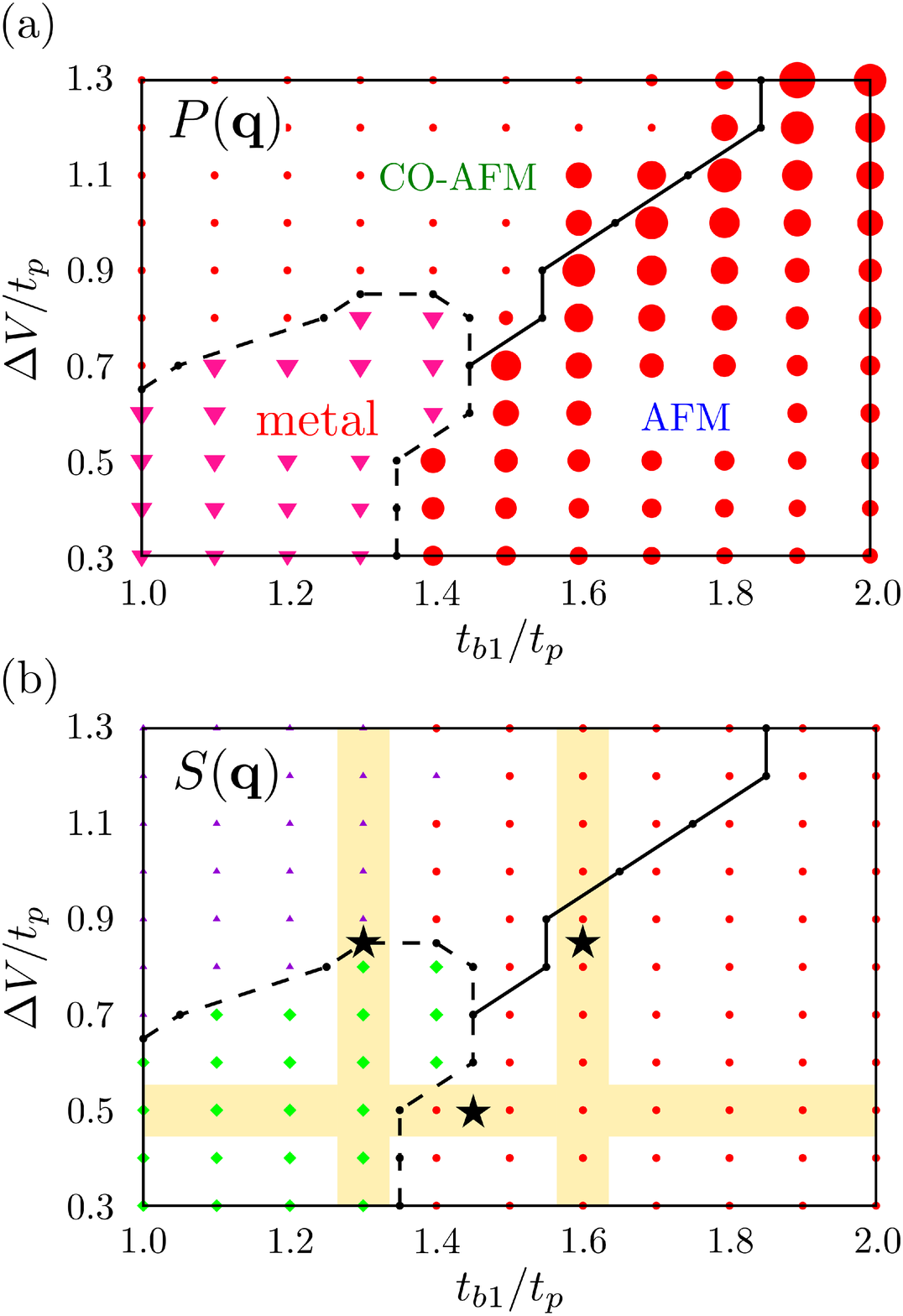}
\end{center}
\caption{(Color online) 
(a) Polar charge correlation function $P({\bf q})$, and (b) spin correlation function $S({\bf q})$ plotted in the $t_{b1}/t_p-\Delta V/t_p$ plane. 
Sizes of the symbols represent the magnitudes of the correlation functions at ${\bf q}_{\rm max}$, where the correlation functions take their maxima.
The smallest size symbols represent the results where the magnitudes are smaller than one. 
Red circles and pink inverted triangles in (a) are for $P(\pi, \pi)$, and $P(2\pi/3, -2\pi/3)$, respectively, 
and green rectangles, red circles, and violet triangles in (b) are for $S(\pi,0)$, $S(\pi,\pi)$, and $S(2\pi/3, 2\pi/3)$, respectively. 
Size scales for the symbols are taken to be the same in (a) and (b) with each other. 
A $6 \times 6$ site cluster is utilized. 
Black stars in (b) represent the phase boundaries calculated in a $12 \times 12$ site cluster. 
}
\label{fig:vtphase}
\end{figure}
The spin correlation function and the polar-charge correlation function are defined as 
\begin{align}
S({\bf q})=4N^{-1} \sum_{i i'} 
\left (\langle S^z_i S^z_{i'} \rangle -\langle S^z_i \rangle \langle S^z_{i'} \rangle \right)
e^{i {\bf q} \cdot ({\bf r}_i-{\bf r}_{i'})} , 
\end{align} 
and
\begin{align}
P({\bf q})=N^{-1} \sum_{ii'} 
\left(\langle P_i P_{i'} \rangle  -\langle P_i \rangle \langle P_{i'} \rangle \right)
e^{i {\bf q} \cdot \left ({\bf r}_i-{\bf r}_{i'}\right)} , 
\end{align} 
respectively.
A prefactor 4 in $S({\bf q})$ is attributed to the 1/2 factor in the spin operator. 
We also introduce the charge correlation function 
\begin{align}
N({\bf q})=N^{-1}\sum_{ii'}
\left (\langle n_{i }n_{i' }\rangle - \langle n_{i }\rangle \langle n_{i' }\rangle \right) e^{i {\bf q}\cdot ({\bf r}_i-{\bf r}_{i'})}, 
\end{align} 
where $n_{i}=\sum_{\mu \sigma} n_{i \mu \sigma}$. 
As explained later, the metallic and insulating phases are identified by the ${\bf q}$ dependence of $N\left( {\bf q} \right)$.

\begin{figure}[t]
\begin{center}
\includegraphics[width=\columnwidth,clip]{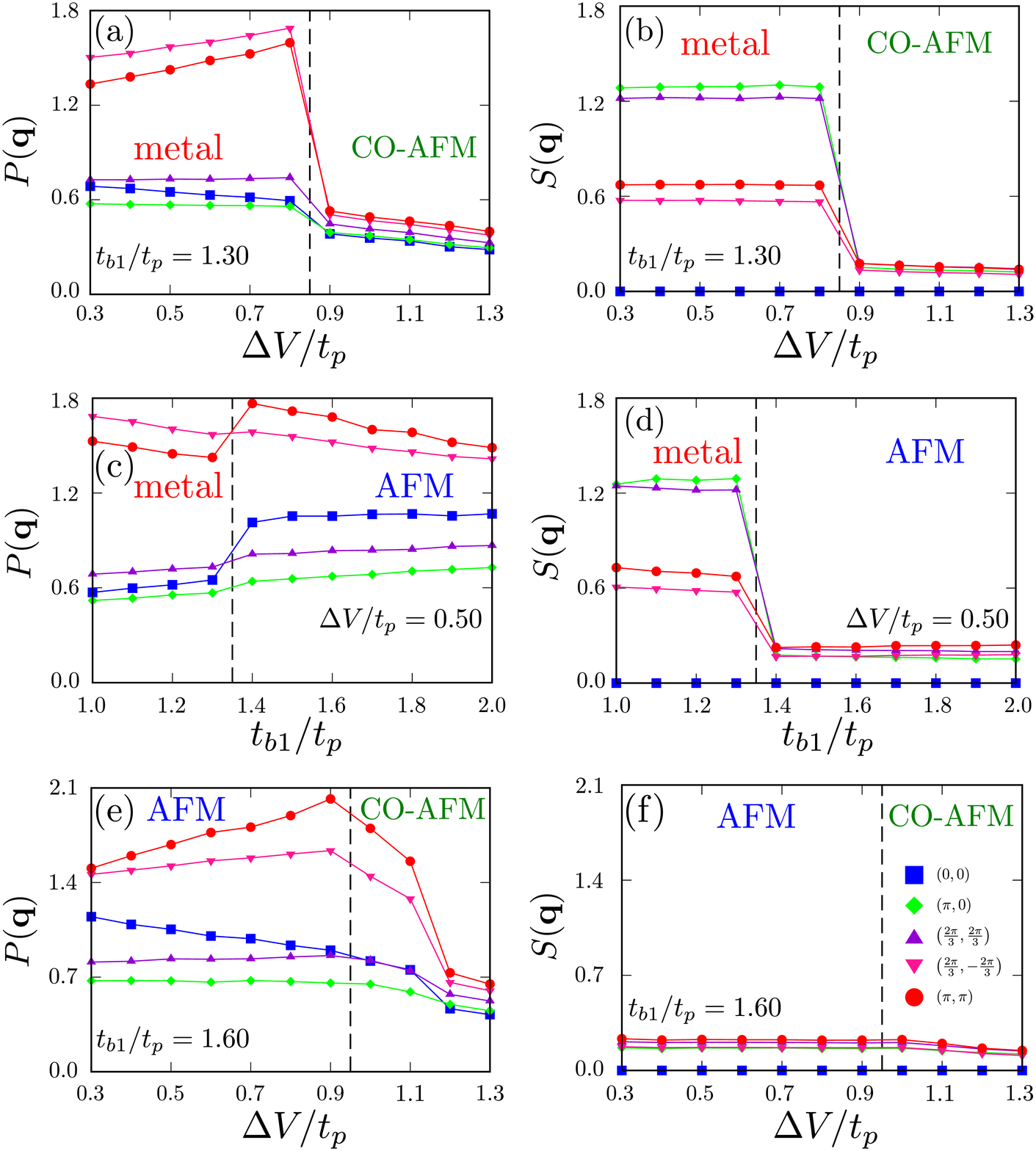}
\end{center}
\caption{(Color online) 
Polar charge correlation functions $P({\bf q})$ and spin correlation functions $S({\bf q})$. 
(a), (b) $\Delta V$ dependence across the metal/CO-AFM phase boundary at $t_{b1}/t_p=1.3$, 
(c), (d) $t_{b1}$ dependence across the metal/AFM phase boundary at $\Delta V/t_p=0.5$, 
and (e), (f) $\Delta V$ dependence across the AFM/CO-AFM boundary at $t_{b1}/t_p=1.6$. 
Adopted parameter values are indicated by the bold lines in Fig.~\ref{fig:vtphase}(b).  
A $6 \times 6$ site cluster is adopted. 
}
\label{fig:correlation}
\end{figure}
%
In Fig.~\ref{fig:vtphase}, we show the phase diagrams in the plane of the intra-dimer hopping $t_{b1}$ and  the inter-dimer Coulomb interaction $\Delta V(\equiv V_p-V_q)$. 
The spin and charge structures are examined in a wide parameter range, although the parameter values for $\kappa$-(BEDT-TTF)$_2$Cu$_2$(CN)$_3$ are approximately at $t_{b1}/t_p=2-3$ and $\Delta V/t_p=0.3$.~\cite{Komatsu,Nakamura,Koretsune} 
Increase of $t_{b1}$ corresponds to the enhancement of the molecular dimerization;  the system is  reduced to the single-band Hubbard model at half filling. 
There are three phases with different spin and charge structures: a metallic phase in a small $\Delta V$ and $t_{b1}$ region, an AFM insulating phase in a large $t_{b1}$ region, and a polar CO insulating phase associated with the AFM order in a large $\Delta V$ region. 
These are termed ``metal", ``AFM", and ``CO-AFM" phases, respectively. 
Sizes of the symbols represent the magnitudes of the correlation functions at ${\bf q}_{\rm max}$, where the  correlation functions take their maxima. 
It is apparently shown that $P({\bf q}_{\rm max})$ is much larger than $S({\bf q}_{\rm max})$ in all parameter region shown in Fig.~\ref{fig:vtphase}. 
In the AFM phase, ${\bf q}_{\rm max}$ for $P({\bf q})$ is $(\pi, \pi)$, corresponding to the charge structure  in the CO-AFM phase. 
On the other hand, in the metal phase, ${\bf q}_{\rm max}=(\pi, 0)$ and $(2\pi/3, -2\pi/3)$ for $S({\bf q})$ and $P({\bf q})$, respectively, which are different from the momenta characterizing the spin and charge structures in the CO-AFM phase. 

Detailed parameter dependences of the correlation functions are presented in Fig.~\ref{fig:correlation}. 
We focus mainly on the results across the MI transitions. 
The $\Delta V$ dependence ($t_{b1}$ dependence) shown in Figs.~\ref{fig:correlation}(a) and \ref{fig:correlation}(b), 
(Figs.~\ref{fig:correlation}(c) and \ref{fig:correlation}(d))  
corresponds to the vertical line (horizontal line) in Fig.~\ref{fig:vtphase}(b). 
The results indicate that both the transitions between the metal/AFM and metal/CO-AFM phases are of the first order. 
As shown in Fig.~\ref{fig:correlation}(a), the dominant polar-CO correlations are $P(2\pi/3, -2\pi/3)$ and $P(\pi, \pi)$. 
With increasing $\Delta V$ in the metal phase, both components increase, and $P(\pi, \pi)$ tends to overtake $P(2\pi/3, -2\pi/3)$. 
All components of $P({\bf q})$ drop discontinuously, when the system is changed into the CO-AFM phase. 
As shown in Fig.~\ref{fig:correlation}(c), small discontinuous changes in $P({\bf q})$ are seen at the metal/AFM phase boundary, and $P(\pi, \pi)$ becomes larger than $P(2\pi/3, -2\pi/3)$ in the AFM phase. 
Reductions of $P({\bf q})$ at $(\pi, \pi)$ and $(2\pi/3, -2\pi/3)$ with increasing $t_{b1}$ in the metal and AFM phases (see Fig.~\ref{fig:correlation}(c)) are attributable  to the enhancement of the dimerization.
In contrast to $P({\bf q})$, all components of $S({\bf q})$ do not show remarkable changes in the metal phase even near the phase boundaries, as shown in Figs.~\ref{fig:correlation}(b) and \ref{fig:correlation}(d). 
This point is consistent with the calculated results in the single band Hubbard model based on the dimer-Mott insulating picture.~\cite{TWatanabe}
Discontinuous reductions of $S({\bf q})$ at the metal/AFM and metal/CO-AFM phase boundaries are due to the emergence of the magnetic orders. 

The correlation functions across the AFM/CO-AFM phase boundary are shown in Figs.~\ref{fig:correlation}(e) and \ref{fig:correlation}(f). 
The changes in the correlation functions are continuous, that is, the second-order phase transition.  
Steep increases and decreases in $P({\bf q})$ around the phase boundary reflect the emergence of the long-range polar CO. Small reduction of $S({\bf q})$ in the CO-AFM phase with increasing $\Delta V$ [see Fig.~\ref{fig:correlation}(f)] implies the competing relation between the polar CO structure and the AFM structure  suggested in the previous studies~\cite{Naka, Hotta}.

\begin{figure}[t]
\begin{center}
\includegraphics[width=\columnwidth,clip]{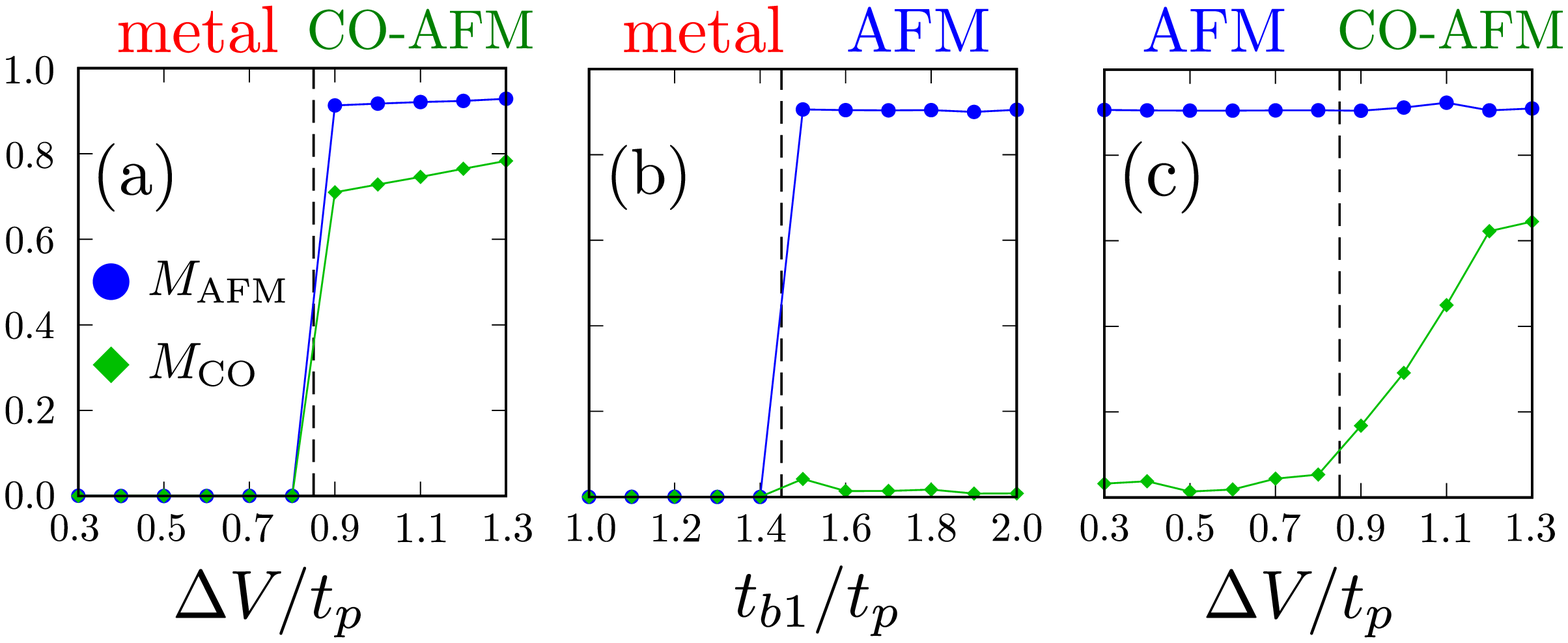}
\end{center}
\caption{(Color online) 
The order parameters for the AFM and polar-CO orders. 
(a) $\Delta V$ dependence across the metal/CO-AFM phase boundary at $t_{b1}/t_p=1.3$, 
(b) $t_{b1}$ dependence across the metal/AFM phase boundary at $\Delta V/t_p=0.5$, 
and (c) $\Delta V$ dependence across the AFM/CO-AFM boundary at $t_{b1}/t_p=1.6$. 
A $12 \times 12$ site cluster is adopted. 
}
\label{fig:oparameter}
\end{figure}

The orders of the phase transition are directly identified by calculating the order parameters. 
In Fig.\ref{fig:oparameter}, $M_{\rm AFM}$ and $M_{\rm CO}$ are plotted along the vertical and horizontal lines in Fig.\ref{fig:vtphase} (b). 
It is shown that both the metal/AFM and metal/CO-AFM transitions (Figs.~\ref{fig:oparameter}(a) and (b)) are of the first order, and the AFM/CO-AFM transition (Fig.~\ref{fig:oparameter}(c)) is of the second one. 
These are consistent with the results in the correlation functions shown in Fig.~\ref{fig:correlation}. 


\begin{figure}[t]
\begin{center}
\includegraphics[width=\columnwidth,clip]{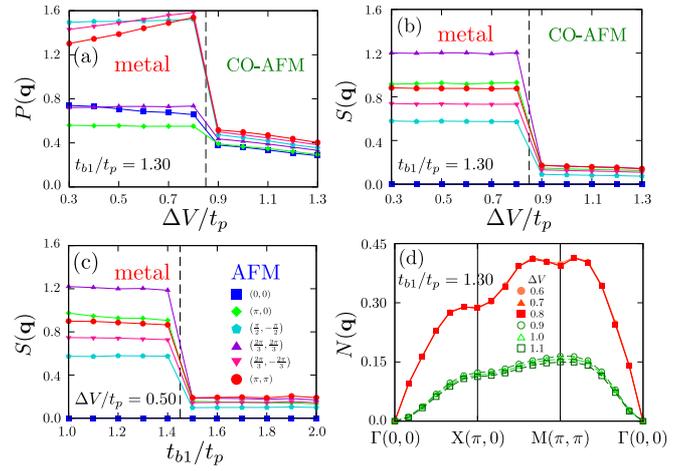}
\end{center}
\caption{(Color online) 
Results in a $12 \times 12$ site cluster. 
(a)-(c) 
Polar charge correlation functions $P({\bf q})$ and spin correlation functions $S({\bf q})$. 
$\Delta V$ dependence across the metal/CO-AFM phase boundary at $t_{b1}/t_p=1.3$ in (a) and (b), and $t_{b1}$ dependence across the metal/AFM phase boundary at $\Delta V/t_p=0.5$ in (c). 
(d) The momentum dependences of the charge correlation functions $N({\bf q})$. 
}
\label{fig:size12}
\end{figure}

To check the size dependences of the numerical results, we examine the spin and charge correlations in the $12 \times 12$-site cluster. 
Calculated results of the phase boundaries are plotted in Fig.~\ref{fig:vtphase}(b), together with the results in the $6 \times 6$-site cluster. 
The results in the two-sizes of clusters are almost coincides with each other.  
The spin and polar-charge correlation functions calculated in the $12\times 12$-site cluster are shown in Figs.~\ref{fig:size12}(a)--(c). 
The polar charge correlation functions at ${\bf q}=(\pi, \pi)$, and $(2\pi/3, -2\pi/3)$ increase toward the metal/CO-AFM phase boundary, and 
$P(\pi, \pi)$ tends to overtake others near the boundary. 
On the other hand, all components of $S({\bf q})$ are almost unchanged even near the metal/AFM phase boundary. 
We confirm that $P({\bf q}_{\rm max})$ are larger than $S({\bf q}_{\rm max})$ in the whole of the calculated parameter sets. 
These are consistent with the results obtained in the $6 \times 6$-site cluster, that is, the size dependence is  not serious. 
The metal and insulating phases are able to be identified by calculating the charge correlation function $N({\bf q})$; the insulating charge gap provides the quadratic ${\bf q}$ dependence in $N({\bf q})$ in the limit of $|{\bf q}| \rightarrow 0$~\cite{Feynman}. 
Figure~\ref{fig:size12}(d) shows that $N({\bf q})$ in a small ${\bf q}$ region changes from the quadratic-like to linear dependences across the metal/CO-AFM phase boundary.  

%
Based on the present results, we discuss the spin and charge structures near the metal/insulator phase boundary, and implications for the $\kappa$-type BEDT-TTF salts. 
First, we confirm that both the metal/AFM insulator transition and the metal/CO-AFM insulator transition are of the first order, and are not associated with the critical spin/charge fluctuations. 
This result suggests that the weak-coupling theories, in which the first-order MI transition due to the electron correlation cannot be treated properly, tend to overestimate the spin and charge fluctuations near the long-range ordered states. 

Second, it is found that the polar-charge correlations are larger than the spin correlations in the metal phase close to the MI phase boundary. 
As summarized in Fig.~\ref{fig:vtphase}, $P({\bf q})$ increase toward the metal/CO-AFM phase boundary. On the other hand, $S({\bf q})$ are almost unchanged even at the vicinity of the metal/AFM phase boundary. 
The results might be attributable to the fact that $P({\bf q})$ near the MI phase boundary are influenced by 
the second-order phase transition between the AFM/CO-AFM phases, where $P({\bf q})$ increases critically. 
As shown in Fig.~\ref{fig:vtphase}, the second-order transition line is terminated at a point, i.e. a critical end point, where this line merges to the first-order transition line, in a similar way to the $\lambda$-line and the gas-liquid transition line in He$^{4}$~\cite{Chaikin}. 
This result suggests that the internal electronic degree of freedom in the molecular dimer is essential for the difference in $P({\bf q})$ and $S({\bf q})$ near the phase boundary, and stimulates us to reinvestigate the mechanism of the superconductivity in $\kappa$-type BEDT-TTF salts. 
So far, we do not mention the Mott insulating phase without the AFM long-range order suggested experimentally because of the limitations of the present theoretical calculations. 
We suppose that, if this phase is realized, this appears at vicinity of the metal/AFM phase boundary, and that the different characters in $S({\bf q})$ and $P({\bf q})$ suggested above are more remarkable in this case.

Third, the spin/charge fluctuations near the carrier-doping induced MI transition are 
qualitatively different from these near the MI transition by the band-width control. 
As well known in HTSC, a carrier doping to the AFM Mott insulator melts the AFM long range order and as a results, a sizable AFM fluctuation remains close to the MI phase boundary. 
On the other hand, the MI transition induced by the band-width control is discontinuous and the critical enhancement of the fluctuations are not expected in the usual manner. 
The present study indicates that, even near the discontinuous phase transition, the electronic degree of freedom in the molecule dimer provides the continuous phase boundary near the MI transition, and enhances the polar-charge fluctuations in a metallic phase.


We thank J.~Nasu, and H.~Seo for their helpful discussions.
This work was supported by MEXT KAKENHI Grant No. 26287070, 15H02100, 16K17731 and 16K17752.
Some of the numerical calculations were performed using the facilities of the Supercomputer Center, the Institute for Solid State Physics, the University of Tokyo.

\end{document}